\documentclass[aps,prb,twocolumn,showpacs,preprintnumbers,amsmath,amssymb, superscriptaddress, bibnotes]{revtex4}
\usepackage{graphicx}
\usepackage{dcolumn}
\usepackage{bm}
\usepackage[colorlinks,linkcolor=blue,anchorcolor=blue,citecolor=blue,urlcolor=blue,filecolor=blue,menucolor=blue,runcolor=blue]{hyperref}
\usepackage{ulem}

\makeatletter
\@ifundefined{textcolor}{}
{%
	\definecolor{BLACK}{gray}{0}
	\definecolor{WHITE}{gray}{1}
	\definecolor{RED}{rgb}{1,0,0}
	\definecolor{GREEN}{rgb}{0,1,0}
	\definecolor{BLUE}{rgb}{0,0,1}
	\definecolor{CYAN}{cmyk}{1,0,0,0}
	\definecolor{MAGENTA}{cmyk}{0,1,0,0}
	\definecolor{YELLOW}{cmyk}{0,0,1,0}
}

\usepackage{color}

\newcommand{\beq}{\begin{eqnarray}}
\newcommand{\eeq}{\end{eqnarray}}
\newcommand{\ys}[1]{\textcolor{black}{#1}}
\newcommand{\bl}[1]{\textcolor{black}{#1}}
\makeatother
\usepackage{babel}

\begin{document}

\title{Strain-sensitive superconductivity in kagome metals KV$_3$Sb$_5$ and CsV$_3$Sb$_5$ probed by point-contact spectroscopy
}
\author{Lichang Yin}
\thanks{These authors have contributed equally to this work.}
\author{Dongting Zhang}
\thanks{These authors have contributed equally to this work.}
\author{Chufan Chen}
\author{Ge Ye}
\affiliation  {Center for Correlated Matter and Department of Physics, Zhejiang University, Hangzhou, 310058, China}
\affiliation  {Zhejiang Province Key Laboratory of Quantum Technology and Device, Department of Physics, Zhejiang University, Hangzhou 310058, China}

\author{Fanghang Yu}
\affiliation  {Hefei National Laboratory for Physical Sciences at Microscale and Department of Physics, and CAS Key Laboratory of Strongly-coupled Quantum Matter Physics, University of Science and Technology of China, Hefei, Anhui 230026, China}

\author{Brenden R. Ortiz}
\affiliation{Materials Department and California Nanosystems Institute, University of California Santa Barbara, Santa Barbara, CA, 93106, United States}
\author{Shuaishuai Luo}
\author{Weiyin Duan}
\author{Hang Su}
\affiliation  {Center for Correlated Matter and Department of Physics, Zhejiang University, Hangzhou, 310058, China}
\affiliation  {Zhejiang Province Key Laboratory of Quantum Technology and Device, Department of Physics, Zhejiang University, Hangzhou 310058, China}
\author{Jianjun Ying}
\affiliation  {Hefei National Laboratory for Physical Sciences at Microscale and Department of Physics, and CAS Key Laboratory of Strongly-coupled Quantum Matter Physics, University of Science and Technology of China, Hefei, Anhui 230026, China}

\author{Stephen D. Wilson}
\affiliation{Materials Department and California Nanosystems Institute, University of California Santa Barbara, Santa Barbara, CA, 93106, United States}
\author{Xianhui Chen}
\affiliation  {Hefei National Laboratory for Physical Sciences at Microscale and Department of Physics, and CAS Key Laboratory of Strongly-coupled Quantum Matter Physics, University of Science and Technology of China, Hefei, Anhui 230026, China}
\affiliation  {CAS Center for Excellence in Quantum Information and Quantum Physics, Hefei, Anhui 230026, China}
\affiliation  {Collaborative Innovation Center of Advanced Microstructures, Nanjing 210093, China}

\author{Huiqiu Yuan}
\affiliation  {Center for Correlated Matter and Department of Physics, Zhejiang University, Hangzhou, 310058, China}
\affiliation  {Zhejiang Province Key Laboratory of Quantum Technology and Device, Department of Physics, Zhejiang University, Hangzhou 310058, China}
\affiliation  {Collaborative Innovation Center of Advanced Microstructures, Nanjing 210093, China}
\affiliation  {State Key Laboratory of Silicon Materials, Zhejiang University, Hangzhou 310058, China}
\author{Yu Song}
\email[Corresponding author: ]{yusong$_$phys@zju.edu.cn}
\affiliation  {Center for Correlated Matter and Department of Physics, Zhejiang University, Hangzhou, 310058, China}
\affiliation  {Zhejiang Province Key Laboratory of Quantum Technology and Device, Department of Physics, Zhejiang University, Hangzhou 310058, China}
\author{Xin Lu}
\email[Corresponding author: ]{xinluphy@zju.edu.cn}
\affiliation  {Center for Correlated Matter and Department of Physics, Zhejiang University, Hangzhou, 310058, China}
\affiliation  {Zhejiang Province Key Laboratory of Quantum Technology and Device, Department of Physics, Zhejiang University, Hangzhou 310058, China}
\affiliation  {Collaborative Innovation Center of Advanced Microstructures, Nanjing 210093, China}
\date{\today}

\begin{abstract}
The kagome lattice is host to flat bands, topological electronic structures, Van Hove singularities and diverse electronic instabilities, providing an ideal platform for realizing highly tunable electronic states. Here, we report soft- and mechanical- point-contact spectroscopy (SPCS and MPCS) studies of the kagome superconductors KV$_3$Sb$_5$ and CsV$_3$Sb$_5$. Compared to the superconducting transition temperature $T_{\rm c}$ from specific heat measurements (2.8~K for CsV$_3$Sb$_5$ and 1.0~K for KV$_3$Sb$_5$), significantly enhanced \ys{values of $T_{\rm c}$ are} observed via the zero-bias conductance of SPCS ($\sim$4.2~K for CsV$_3$Sb$_5$ and $\sim$1.8~K for KV$_3$Sb$_5$), which become further enhanced in MPCS measurements ($\sim$5.0~K for CsV$_3$Sb$_5$ and $\sim$3.1~K for KV$_3$Sb$_5$). \ys{While} the differential conductance curves from SPCS \ys{are} described by a two-gap $s$-wave model, a single $s$-wave gap \ys{reasonably captures} the MPCS data, \ys{likely due to a} diminishing spectral weight of the other gap. The enhanced superconductivity \bl{probably} \ys{arises} from local strain caused by the point-contact, \ys{which also leads to the evolution from two-gap to single-gap behaviors in different point-contacts.}
Our results demonstrate highly strain-sensitive superconductivity in kagome metals CsV$_3$Sb$_5$ and KV$_3$Sb$_5$, which may be harnessed in the manipulation of possible Majorana zero modes.
\end{abstract}

\maketitle

\section{Introduction}

Due to its unique geometry, the kagome lattice natively hosts electronic flat bands, Dirac band crossings, and Van Hove singularities, allowing for the realization of distinct topological electronic states \cite{Ye2018,Liu2018,Kang2019,Yin2020} and correlated collective orders \cite{Wang2013,Isakov2006,Guo2009,Kiesel2013,Wen2010,park2021electronic}. The recent discovery of superconductivity in the kagome metals $A$V$_3$Sb$_5$ ($A=$~K, Rb, Cs) \cite{Ortiz2019,Ortiz2020,ortiz2020superconductivity,yin2021superconductivity} triggered immense interest, as superconductivity in these materials coexist with topologically protected surface states \cite{Ortiz2020} and an unusual chiral charge order \cite{jiang2020discovery,Feng_2021}, offering an ideal platform to investigate the interplay between these exotic phenomena and their evolution upon tuning.

While the nature of superconducting pairing in $A$V$_3$Sb$_5$ is still under debate \cite{zhao2021nodal,duan2021nodeless,mu2021swave,xu2021multiband,wu2021nature}, signatures of spin-triplet supercurrent were found in K$_{1-x}$V$_3$Sb$_5$ \ys{Josephson junctions} \cite{wang2020proximityinduced}, and possible Majorana zero modes have been detected in CsV$_3$Sb$_5$ \cite{liang2021threedimensional}. These findings raise the possibility that $A$V$_3$Sb$_5$ may exhibit topological superconductivity with \ys{potential} applications in fault-tolerant quantum computation \cite{Sarma2015,TSCreview1}. Superconductivity in $A$V$_3$Sb$_5$ is highly susceptible to pressure, displaying two superconducting domes in the temperature-pressure phase diagram \cite{zhao2021nodal, chen2021double, du2021pressuretuned, Yu2021, zhang2021pressureinduced, zhu2021doubledome}, and a roughly triple enhancement of the superconducting transition temperature $T_{\rm c}$ can be realized under modest pressures of $\approx1$~GPa. The tunability of superconductivity under pressure suggests \ys{that it may be modulated by strain, for example}
to induce superconductor-metal transitions or to stabilize superconductor-metal heterostructures, and \ys{raises} prospects {for the} strain-manipulation of possible Majorana zero modes.

In this work, we applied soft and mechanical point-contact spectroscopy (SPCS and MPCS) to investigate the superconducting properties of single crystalline CsV$_3$Sb$_5$ and KV$_3$Sb$_5$. From both temperature- and field-dependence of the zero-bias conductance, as well as analyses of the differential conductance curves $G$($V$) \ys{with} the Blonder-Tinkham-Klapwijk (BTK) model, we observed \ys{that values of} $T_{\rm c}$ in SPCS and MPCS are substantially enhanced relative to those from \ys{thermodynamic} measurements. The enhancement of superconductivity is attributed to local strain in the point-contact region, \ys{consistent with} the larger enhancement observed in MPCS. While \ys{describing} the differential conductance from SPCS requires two $s$-wave gaps, a single gap is sufficient to \ys{capture the} MPCS results. Nonetheless, the anomalously small ratio between the superconducting gap and $T_{\rm c}$ in the MPCS results suggests the presence of an undetected larger gap.
Our results demonstrate highly strain-sensitive superconductivity in the kagome metals KV$_3$Sb$_5$ and CsV$_3$Sb$_5$, and provide evidence for nodeless multi-gap pairing.

\section{Experimental details}

High quality single crystals of CsV$_3$Sb$_5$ and KV$_3$Sb$_5$ were synthesized using the self-flux method \cite{Ortiz2020,ortiz2020superconductivity}. SPCS measurements were performed by attaching a gold wire (30~$\mu$m in diameter) onto cleaved samples through a drop of silver paint, forming hundreds of parallel nanoscale conducting channels between individual silver particles and the sample surface. \ys{The contact areas have diameters in the range $50 - 100$~$\mu$m}. MPCS measurements were carried out in an anvil-needle configuration, where electrochemically etched gold tips are employed and piezo-controlled nano-positioners are used to gently control the engagement between \ys{the} tip and \ys{the} sample. Differential conductance as a function of the bias voltage $G$($V$) was recorded with the lock-in technique in a quasi-four-probe configuration. An Oxford Instruments cryostat with a $^3$He insert (base temperature 0.3~K) was used for SPCS and MPCS measurements, and magnetic fields up to 3.5~T were applied along the $c$-axis.

\begin{figure}
	\includegraphics[angle=0,width=0.49\textwidth]{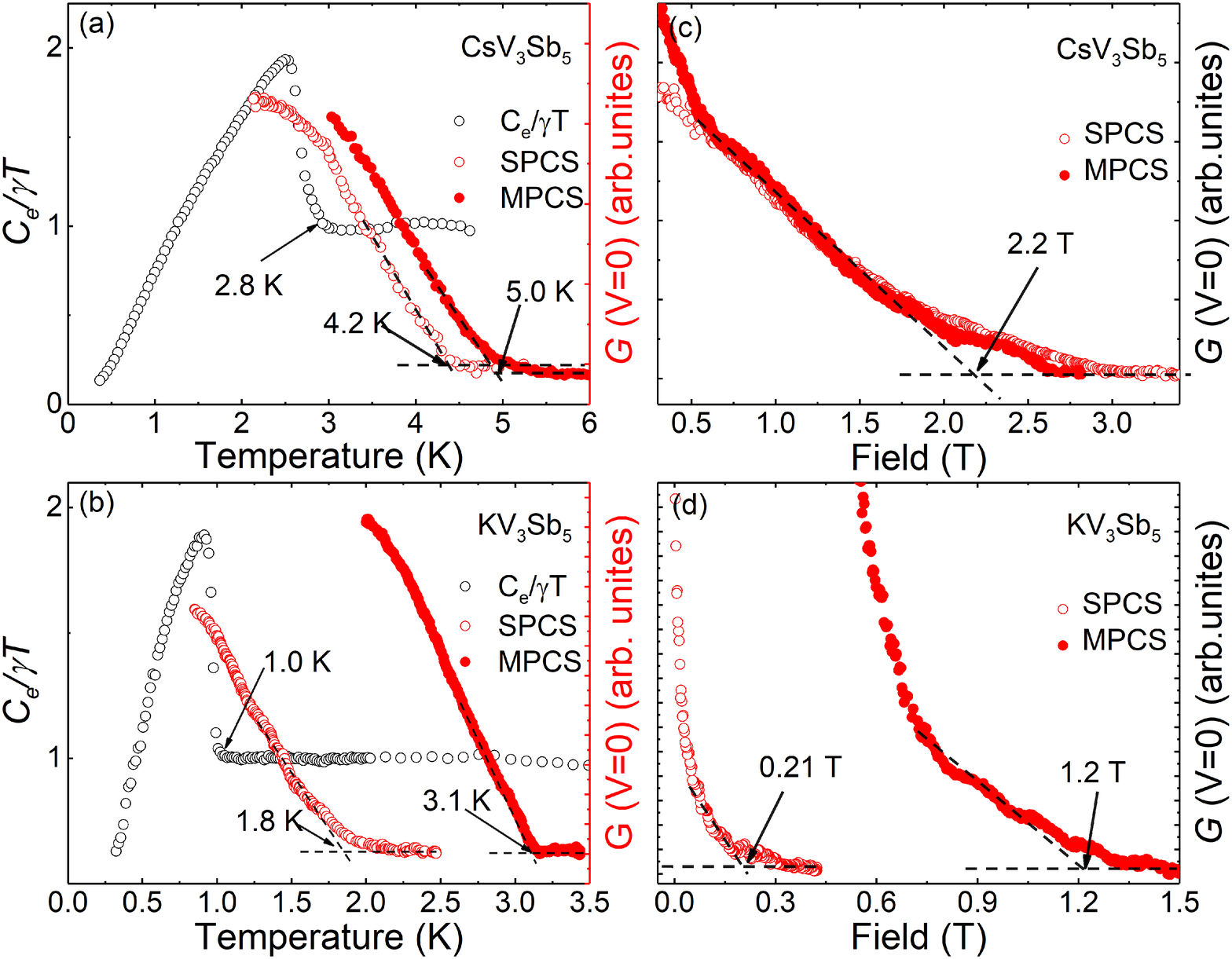}
	\vspace{-12pt} \caption{\label{Figure0} Temperature dependence of the electronic specific heat $C_{\rm e}(T)/\gamma T$ and zero-bias conductance $G$($V$=0) measured using SPCS and MPCS for (a) CsV$_3$Sb$_5$ and (b) KV$_3$Sb$_5$. $C_{\rm e}(T)/\gamma T$ data for CsV$_3$Sb$_5$ and KV$_3$Sb$_5$ are respectively from Ref.~\cite{duan2021nodeless} and Ref.~\cite{du2021pressuretuned}. Field-dependence of the zero-bias conductance $G$($V$=0) measured using SPCS and MPCS at $T=0.3$~K for (c) CsV$_3$Sb$_5$ and (d) KV$_3$Sb$_5$, with the field along the $c$-axis. The arrows mark the onset of superconductivity.
	}
	\vspace{-12pt}
\end{figure}

\section{Results}
\subsection{Zero-bias conductance}

In Fig.~\ref{Figure0}(a), the zero-bias conductance $G$($V$=0) from SPCS and MPCS measurements are compared with the electronic specific heat $C_{\rm e}(T)/\gamma T$ for CsV$_3$Sb$_5$. In contrast to superconductivity that onsets below $T_{\rm c}\approx2.8$~K in $C_{\rm e}(T)/\gamma T$, the temperature dependence of $G$($V$=0) for CsV$_3$Sb$_5$ indicates an onset of Andreev reflection below $\approx4.2$~K for SPCS and $\approx5.0$~K for MPCS, suggesting the $T_{\rm c}$ probed by point-contact spectroscopy \ys{is} significantly enhanced. Similar behaviors are observed in KV$_3$Sb$_5$, with a $T_{\rm c}\approx1.0$~K in $C_{\rm e}(T)/\gamma T$ \ys{increased} to $\approx1.8$~K for SPCS and $\approx3.1$~K for MPCS, as shown in Fig.~\ref{Figure0}(b). \ys{It should be emphasized that while signatures of superconductivity from both Andreev reflection [Figs.~\ref{Figure0}(a) and (b)] and resistivity onset at higher temperatures relative to $C_{\rm e}(T)/\gamma T$  measurements \cite{Ortiz2020,ortiz2020superconductivity}, they are distinct because (1) the onset temperature from Andreev reflection is even higher than that from resistivity, and (2) it increases systematically from SPCS to MPCS in both CsV$_3$Sb$_5$ and KV$_3$Sb$_5$.}

\ys{Instead, given that} enhanced superconductivity is not observed in scanning tunneling microscopy measurements \cite{liang2021threedimensional, Cs-STM2}, the increase of $T_{\rm c}$ from Andreev reflection in Figs.~\ref{Figure0}(a)-(b) likely results from effects of point-contacts on the sample. \bl{Since} superconductivity in $A$V$_3$Sb$_5$ is highly responsive to pressure \cite{zhao2021nodal,chen2021double,du2021pressuretuned,Yu2021,zhang2021pressureinduced,zhu2021doubledome}, local strain induced by point-contacts \ys{may be responsible for the} enhanced superconductivity. The observation of a larger tuning effect in MPCS relative to SPCS is consistent with this scenario, since mechanical point-contacts \ys{typically lead to} a larger strain compared to \ys{soft point-contacts}. The \ys{enhanced} superconductivity also \ys{manifests through} increased upper critical fields $H_{\rm c2}$ at $T$~=~0.3~K ($H\parallel c$): \ys{it is $\approx$~0.21~T for SPCS and $\approx$~1.2~T for MPCS measurements on KV$_3$Sb$_5$ [Fig.~\ref{Figure0}(d)]. In CsV$_3$Sb$_5$, we find $H_{\rm c2}\approx$~2.2~T ($H\parallel c$) for both SPCS and MPCS [Fig.~\ref{Figure0}(c)], significantly higher than $H_{\rm c2}\approx$~1.0~T determined from resistivity measurements \cite{Ni2021}.}

\begin{figure}
	\includegraphics[angle=0,width=0.49\textwidth]{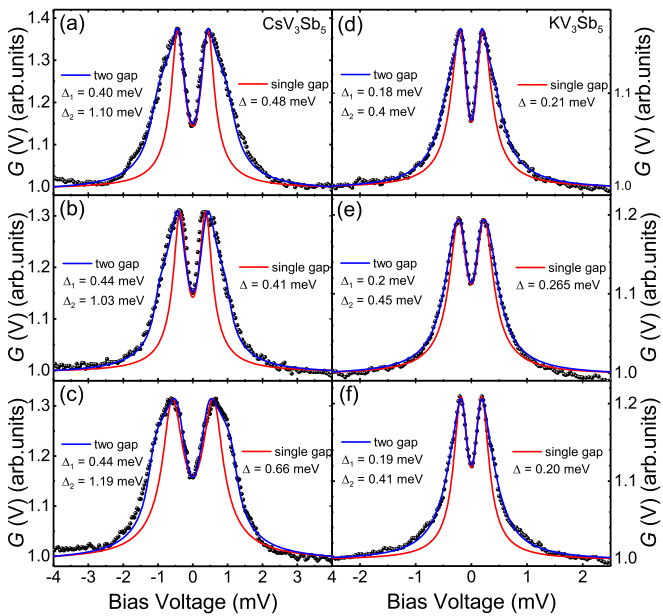}
	\vspace{-12pt} \caption{\label{Figure1} Representative normalized SPCS differential conductance curves for (a)-(c) CsV$_3$Sb$_5$ and (d)-(f) KV$_3$Sb$_5$ at $T$~=~0.3~K. \ys{Solid red lines are fits to a single-gap $s$-wave model for data with voltages below the peak in $G$($V$).} The solid blue lines are fits to a two-gap $s$-wave model. The slight deviation of the measured differential conductance from the fits at large bias voltages is due to current heating effects \cite{PhysRevB.69.134507}.}
	\vspace{-12pt}
\end{figure}

\subsection{Differential conductance curves from SPCS}

\begin{figure}
\includegraphics[angle=0,width=0.49\textwidth]{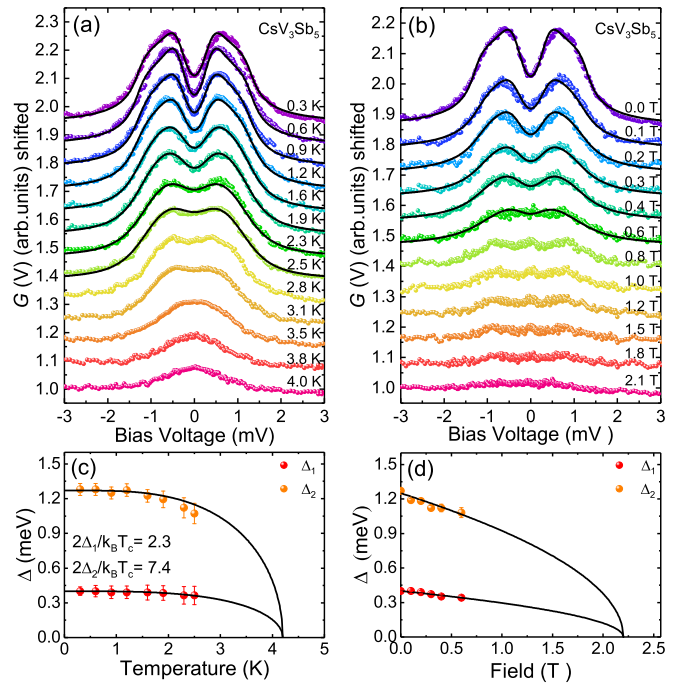}
\vspace{-12pt} \caption{\label{Figure2} Normalized differential conductance curves for CsV$_3$Sb$_5$ from SPCS, as a function of (a) temperature and (b) magnetic field. The solid lines are fits to the two-gap $s$-wave BTK model. The extracted superconducting gaps $\Delta_{1}$ and $\Delta_{2}$ are shown as a function of (c) temperature with $H=0$~T and (d) magnetic field with $T=0.3$~K. The solid lines are based on the BCS theory, with $T_{\rm c}$ and $H_{\rm c2}$ estimated from the zero-bias conductance.
}
\vspace{-12pt}
\end{figure}

\begin{figure}
\includegraphics[angle=0,width=0.49\textwidth]{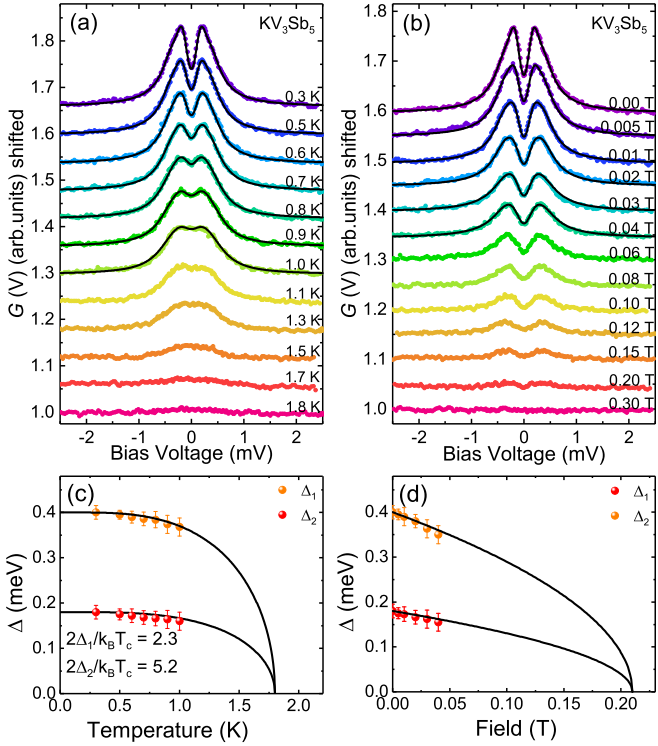}
\vspace{-12pt} \caption{\label{Figure3} Normalized differential conductance curves for KV$_3$Sb$_5$ from SPCS, as a function of (a) temperature and (b) magnetic field. The solid lines are fits to the two-gap $s$-wave BTK model. The extracted superconducting gaps $\Delta_{1}$ and $\Delta_{2}$ are shown as a function of (c) temperature with $H=0$~T and (d) magnetic field with $T=0.3$~K. The solid lines are based on the BCS theory, with $T_{\rm c}$ and $H_{\rm c2}$ estimated from the zero-bias conductance.}
\vspace{-12pt}
\end{figure}

\begin{figure}
\includegraphics[angle=0,width=0.49\textwidth]{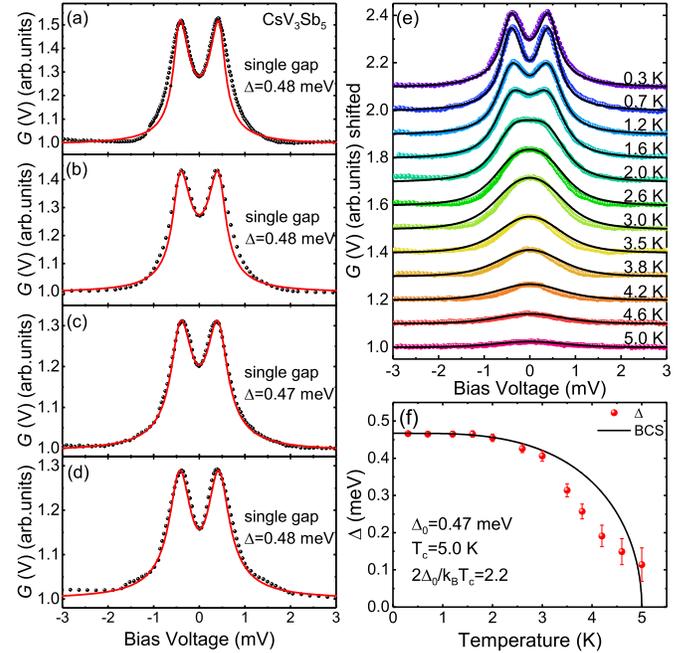}
\vspace{-12pt} \caption{\label{Figure4} (a)-(d) Representative MPCS differential conductance curves for CsV$_3$Sb$_5$ at $T=0.3$~K. The red solid line are fits to a single-gap $s$-wave BTK model. (e) Temperature evolution of the MPCS differential conductance curves for CsV$_3$Sb$_5$, fit to a single-gap $s$-wave BTK model (black solid lines). (f) Temperature dependence of the extracted superconducting gap $\Delta$ from (e), the black line is the expected behavior of the BCS theory, with $\Delta_{0}=0.47$~meV and $T_{\rm c}=5.0$~K ($2\Delta_{0}/k\rm_{B}T\rm_{c}$~=~2.2).}
\vspace{-12pt}
\end{figure}

\begin{figure}
\includegraphics[angle=0,width=0.49\textwidth]{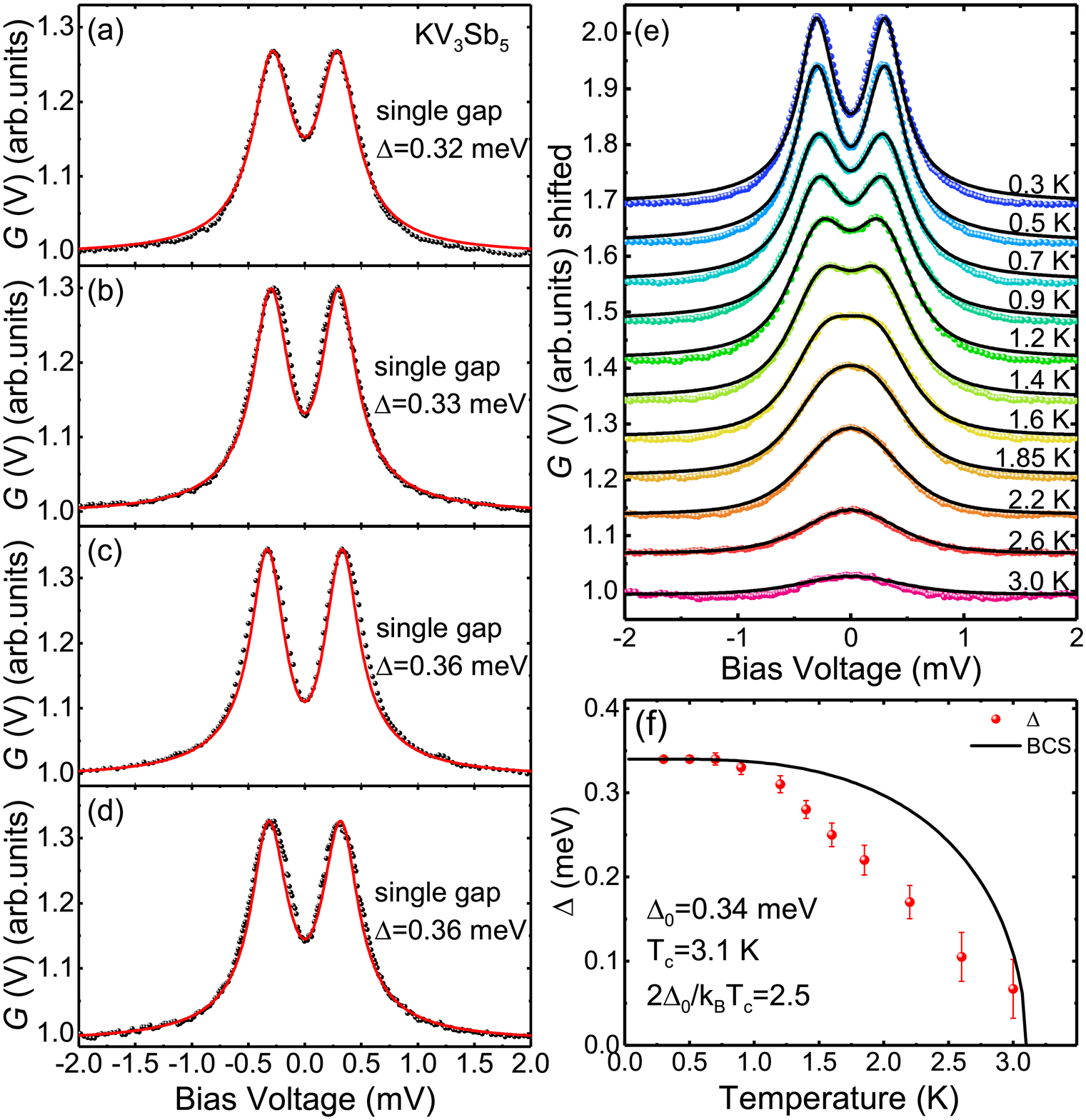}
\vspace{-12pt} \caption{\label{Figure5} (a)-(d) Representative MPCS differential conductance curves for KV$_3$Sb$_5$ at $T$~=~0.3~K. The solid red lines are fits to a single-gap $s$-wave BTK model. (e) Temperature evolution of the MPCS differential conductance curves for KV$_3$Sb$_5$, fit to a single-gap $s$-wave BTK model (black solid lines). (f) Temperature dependence of the extracted superconducting gap $\Delta$ from (e), the black line is the expected behavior of the BCS theory, with $\Delta_{0}=0.34$~meV and $T_{\rm c}=3.1$~K ($2\Delta_{0}/k\rm_{B}T\rm_{c}$~=~2.5).}
\vspace{-12pt}
\end{figure}

To probe the superconducting state with enhanced $T_{\rm c}$, we systematically measured the differential conductance curves $G$($V$) for CsV$_3$Sb$_5$ and KV$_3$Sb$_5$ at various temperatures and under different magnetic fields, with SPCS results in Figs.~\ref{Figure1}-\ref{Figure3}, and MPCS results in Figs.~\ref{Figure4}-\ref{Figure5}.
A representative set of SPCS $G$($V$) curves at 0.3~K for CsV$_3$Sb$_5$ are shown in Figs.~\ref{Figure1}(a)-(c), while those for KV$_3$Sb$_5$ are shown in Figs.~\ref{Figure1}(d)-(f). All the conductance curves for CsV$_3$Sb$_5$ show a double-peak feature around 0.5~mV and a small bulge around 1~mV, characteristic of two-gap superconductivity. A single-gap $s$-wave BTK model fails to describe the $G$($V$) curves, \ys{with substantial deviations of the fits from experimental data}. On the other hand, a two-band BTK \ys{model}, $G(V) = \omega G_{1}(V) + (1-\omega) G_{2}(V)$ ($0\leq\omega\leq1$), with a small ($\Delta_1\sim$0.4 meV) and a large gap ($\Delta_2\sim$1.1 meV) \ys{captures the experimental data, shown as solid blue lines in Figs.~\ref{Figure1}(a)-(c)}. $\omega$ ranges from 60$\%$-80$\%$, indicating that the small gap exhibits a dominant spectral weight.
\ys{Similarly, we find that the differential conductance $G$($V$) curves for KV$_3$Sb$_5$ from SPCS [solid red line in Figs.~\ref{Figure1}(d)-(f)] cannot be satisfactorily captured by a single-gap $s$-wave model. On the other hand, the data can be consistently described by a twp-gap $s$-wave model with $\Delta_1\sim0.18$~meV and $\Delta_2\sim0.38$~meV, with $\omega$ in the range of 20$\%$-70$\%$.}
Our results \ys{suggest that} \ys{despite enhanced gap values induced by local strain in our SPCS measurements, the }superconducting gap structures in both CsV$_3$Sb$_5$ and KV$_3$Sb$_5$ can be appropriately described by a two-gap $s$-wave model, \ys{consistent with nodeless superconductivity in CsV$_3$Sb$_5$ revealed through magnetic penetration depth measurements \cite{duan2021nodeless}}.

Temperature evolution of the differential conductance $G$($V$) curves for SPCS on CsV$_3$Sb$_5$ is shown in Fig.~\ref{Figure2}(a). With increasing temperature, the double peaks gradually shift towards the center, merging into a single zero-bias peak that disappears as temperature approaches $T_{\rm c}$. To extract \ys{the} temperature dependence of the superconducting gaps $\Delta_1$ and $\Delta_2$, we fit the $G$($V$) curves to the two-gap $s$-wave BTK model with $\omega$ constrained to its value at 0.3~K ($\omega=0.674$), shown as solid lines in Fig.~\ref{Figure2}(a). While the two-gap model is clearly better than the single-gap model \ys{in describing} $G(V)$ curves at low temperatures (Fig.~\ref{Figure1}), for $T\gtrsim2.5$~K \ys{both models can reasonably capture the data}. In such a situation, although a two-gap behavior is expected to persist, it is no longer possible to reliably and  independently determine $\Delta_1$ and $\Delta_2$. The extracted \ys{values of} $\Delta_1$ and $\Delta_2$ from \ys{fits} in Fig.~\ref{Figure2}(a) are shown in Fig.~\ref{Figure2}(c), and the solid black lines are the expected behavior from BCS theory, with $T_{\rm c}=4.2$~K inferred from the zero-bias conductance in Fig.~\ref{Figure0}(a). The superconducting gaps at zero temperature are found to be $\Delta_1=0.4$~meV and $\Delta_2=1.27$~meV, yielding $2\Delta_1/k_{\rm B}T_{\rm c}=2.3$ and $2\Delta_2/k_{\rm B}T_{\rm c}=7.4$. The larger gap clearly exceeds $2\Delta/k_{\rm B}T_{\rm c}=3.52$ in the weak-coupling limit, while the smaller gap is well below it. Such a behavior is characteristic of two-gap superconductors \ys{such as} MgB$_2$ \cite{twoband}, and corroborates the notion that superconductivity in CsV$_3$Sb$_5$ is multi-gap \ys{in} nature.

The magnetic field dependence of $G$($V$) for SPCS on CsV$_3$Sb$_5$ at 0.3 K is shown in Fig.~\ref{Figure2}(\bl{b}), with field along the $c$-axis. These data are analyzed \ys{using} the two-gap BTK model, with results shown in Fig.~\ref{Figure2}(d). For $H>0.6$~T, it is no longer possible to independently extract both gaps. The obtained field evolution of the two gaps $\Delta_1$ and $\Delta_2$ reasonably follow $\Delta=\Delta(0)(1-H/H_{\rm c2})^{1/2}$ (solid black lines), \ys{expected} for type-II superconductors in the vortex state \cite{Hc2}. The upper critical field $H_{\rm c2}$ is estimated to be $\approx2.2$~T \ys{from the field-dependence of the} zero-bias conductance [Fig.~\ref{Figure0}(c)].

Similar SPCS measurements were carried out for KV$_3$Sb$_5$, \ys{with} results shown in Fig.~\ref{Figure3}. \ys{The} differential conductance curves \ys{at different temperatures} show that they flatten above $\approx1.8$~K [Fig.~\ref{Figure3}(a)], significantly above $T_{\rm c}=1.0$~K from specific heat.
\ys{For KV$_3$Sb$_5$, the values of $\Delta_1$ and $\Delta_2$ can be reliably extracted for $T\leq1.0$~K ($H$~=~0~T) and $H\leq0.04$~T ($T$~=~0.3~K), with extracted values} \bl{shown in Figs.~\ref{Figure3}(c) and (d), respectively.} The temperature and field evolution of the two gaps are consistent with \ys{expectations of the BCS theory}, with $T_{\rm c}$~=~1.8~K and $H_{\rm c2}$~=~0.21~T determined from the zero-bias conductance [Figs.~\ref{Figure0}(b) and (d)]. \ys{From the extracted gap values, we find $2\Delta_1/k_{\rm B}T_{\rm c}=2.3$ and $2\Delta_2/k_{\rm B}T_{\rm c}=5.2$, similar to CsV$_3$Sb$_5$.}

\subsection{Differential conductance curves from MPCS}

In order to have a comparative study \ys{of how strain affects} the superconducting state of $A$V$_3$Sb$_5$, we applied MPCS to study both CsV$_3$Sb$_5$ and KV$_3$Sb$_5$, with results shown in Fig.~\ref{Figure4} for CsV$_3$Sb$_5$ and Fig.~\ref{Figure5} for KV$_3$Sb$_5$.
In the case of CsV$_3$Sb$_3$, we find the measured differential conductance curves can be reasonably fit by a single-gap $s$-wave model [Figs.~
\ref{Figure5}(a)-(d)], with $\Delta\approx0.47$~meV. However, from the zero-bias conductance [Fig.~\ref{Figure0}(a)] and temperature-dependent conductance curves [Fig.~\ref{Figure4}(e)], we find $T_{\rm c}\approx5.0$~K, which leads to $2\Delta/k_{\rm B}T_{\rm c}\approx2.2$. This value is significantly smaller than the BCS weak-coupling limit of 3.52, but is similar to $2\Delta_1/k_{\rm B}T_{\rm c}\approx2.3$ from SPCS measurements \bl{on CsV$_3$Sb$_5$}. This implies the presence of a larger gap in MPCS measurements, which is \ys{undetected} due to a diminishing spectral weight. By analyzing the differential conductance curves with the single-gap $s$-wave model, we find \ys{the} temperature dependence of $\Delta$ clearly deviates from BCS theory [Fig.~\ref{Figure4}(f)], \ys{characteristic of} the smaller gap in some two-gap superconductors \cite{twogapratio1,twogapratio2}.

Similar behaviors are found for MPCS on KV$_3$Sb$_5$, as shown in Fig.~\ref{Figure5}. The differential conductance \ys{curves are also} reasonably accounted for using a single-gap $s$-wave model, with $\Delta\approx0.34$~meV [Figs.~\ref{Figure5}(a)-(d)]. Combined with $T_{\rm c}\approx3.1$~K from Figs.~\ref{Figure0}(b) and \ref{Figure5}(e), we \ys{obtain} $2\Delta/k_{\rm B}T_{\rm c}\approx2.5$, also significantly smaller than 3.52, \bl{and clear deviations of $\Delta$ from the BCS theory is also observed in its temperature evolution.} Our MPCS measurements suggest that CsV$_3$Sb$_5$ and KV$_3$Sb$_5$ are similar, and both exhibit multi-gap superconductivity, although the larger gap is \ys{undetected likely} due to a diminished spectral weight.

\section{Discussion and conclusion}

All differential conductance curves in Figs.~\ref{Figure1}-\ref{Figure5} from SPCS and MPCS can be \ys{consistently }described by a two-gap $s$-wave model, although extracting \ys{the }values of the two gaps may be hindered by the diminishing spectral weight of the larger gap in MPCS measurements. Our results therefore favor nodeless multi-gap superconductivity in the $A$V$_3$Sb$_5$ series, and suggest \ys{that the structure of the} superconducting order parameter remains robust when $T_{\rm c}$ becomes strongly enhanced. This is consistent with the observation of an exponential temperature dependence of the magnetic penetration depth \cite{duan2021nodeless}, a Hebel-Slichter coherence peak from nuclear magnetic resonance \cite{mu2021swave}, and sensitivity of the superconducting state to magnetic impurities from scanning tunneling microscopy measurements \cite{xu2021multiband}. However, it should be noted that a robust nodal superconducting state \ys{has been suggested by} thermal conductivity measurements \cite{zhao2021nodal}, and supported by \ys{certain} theoretical considerations \cite{wu2021nature}.

In point-contact spectroscopy measurements on layered materials, perfectly two-dimensional Fermi sheets do not contribute to the differential conductance for point-contact in \ys{the} $c$-axis direction. As a Fermi surface becomes progressively more corrugated along the $c$-axis direction, its spectral weight in the differential conductance increases in tandem. Therefore, the differing spectral weight of the large and small superconducting gaps
between SPCS and MPCS measurements may correspond to varying degrees of out-of-plane dispersion in the normal state electronic structures. \ys{In this scenario, the small spectral weight of the larger gap in MPCS measurements suggests that the larger strain in MPCS measurements leads to the corresponding band becoming more two-dimensional compared to SPCS measurements, and points to a nontrivial effect of strain on the normal state electronic structure of $A$V$_3$Sb$_5$.}

For SPCS measurements, local stress or strain may arise during curing of the silver paste or upon cooling due to differential thermal contraction between the sample and the silver paste. As these effects are usually weak, they can be ignored in most SPCS measurements. However, given the tendency \ys{of $A$V$_3$Sb$_5$ crystals} to exfoliate, lateral stress \ys{could be} limited to a few layers near \ys{the} surface, which can induce a sizable strain even if the stress is small. \ys{Such sizable strains coupled with the} sensitivity to pressure (and thus strain) \cite{zhao2021nodal,chen2021double,du2021pressuretuned,Yu2021,zhang2021pressureinduced,zhu2021doubledome}, likely accounts for the enhanced superconductivity observed in CsV$_3$Sb$_5$ and KV$_3$Sb$_5$ from our SPCS measurements. In addition, as $A$V$_3$Sb$_5$ are good metals with a large density of states at the Fermi level, charge doping through the silver paste should play a minor role in enhancing superconductivity. In contrast,
\ys{both strain effects and charge doping may be operative in the dramatic enhancement of $T_{\rm c}$ seen in MoTe$_2$ from SPCS measurements \cite{MoTe2}, as MoTe$_2$ is a semi-metal with superconductivity sensitive to pressure \cite{Qi2016}}.

Compared to SPCS, a more significant strain is typically applied by the sharp tip in MPCS measurements, consistent with the more \ys{significant} increase in $T_{\rm c}$. It should be noted that whereas strain is dominantly in the $ab$-plane for SPCS, it is mostly along the $c$-axis for MPCS. The observation of enhanced $T_{\rm c}$ in both approaches then indicates superconductivity in $A$V$_3$Sb$_5$ is sensitive to strain in more than one symmetry channel, and motivates further studies on the evolution of superconductivity in $A$V$_3$Sb$_5$ under uniaxial strain. \ys{In addition, since the $c$-axis collapses more readily under hydrostatic pressure \cite{zhang2021pressureinduced}, the large susceptibility of the crystal structure to $c$-axis stress may also contribute to the stronger effect  of mechanical point-contact on superconductivity.} A dramatic increase in $T_{\rm c}$ was also reported in FeSe through MPCS measurements \cite{FeSe}, which is also a highly two-dimensional \ys{material with superconductivity that becomes enhanced} under hydrostatic pressure \cite{Medvedev2009,Margadonna2009}. \ys{Taking the  results on $A$V$_3$Sb$_5$, MoTe$_2$ and FeSe together, it appears that layered materials with \ys{superconductivity that becomes enhanced} under hydrostatic pressure are susceptible to strain effects in point-contact spectroscopy measurements.}

\ys{Compared to hydrostatic pressure, strain is much easier to realize and manipulate in devices, giving it a unique advantage in practical applications.} Our results demonstrate that local strain from point-contacts are sufficient to dramatically tune the superconducting properties of $A$V$_3$Sb$_5$, and suggests that such strain-sensitive superconductivity may be useful in device engineering, for example, in realizing superconductor-metal heterostructures. Furthermore, strain may be utilized in the manipulation of possible Majorana zero modes in these materials \cite{liang2021threedimensional}.

In conclusion, we performed \ys{SPCS and MPCS measurements} on CsV$_3$Sb$_5$ and KV$_3$Sb$_5$, and observed enhanced superconductivity in all cases. This enhancement likely originates from local strain induced by the point-contacts, since similar enhancements are absent in scanning tunneling microscopy measurements. While the superconducting gap \ys{structure is} better described by a two-gap $s$-wave model for SPCS results (as in MgB$_2$), a single-gap BTK model can reasonably capture the MPCS results, pointing to a nontrivial evolution of the normal state electronic structure upon strain-tuning.
Our findings highlight the sensitivity of superconductivity to strain in the kagome metals $A$V$_3$Sb$_5$, which may be useful in \ys{the device-engineering of these materials.}

We are grateful for valuable discussions with C. Cao and Y. Liu. The work at Zhejiang University was supported by the National Key R\&D Program of China (No. 2017YFA0303100, No. 2016YFA0300202), the Key R\&D Program of Zhejiang Province, China (2021C01002), the National Natural Science Foundation of China (No. 11674279, No. 11974306 and No. 12034017), and the Fundamental Research Funds for the Central Universities of China. X.L. would like to acknowledge support from the Zhejiang Provincial Natural Science Foundation of China (LR18A04001). S.D.W. and B.R.O. gratefully acknowledge support via the UC Santa Barbara NSF Quantum Foundry funded via the Q-AMASE-i program under award DMR-1906325. B.R.O. also acknowledges support from the California NanoSystems Institute through the Elings fellowship program.

\bibliography{CsV3Sb5RefV3}

\end{document}